\documentclass[english,pre,10pt]{revtex4}
\usepackage{amsfonts}
\usepackage{amsmath}
\usepackage{graphicx}
\newcommand{\norm}[1]{\left|\left| #1\right|\right|}
\usepackage{color}
\usepackage{babel}
\usepackage{hyperref}
\usepackage{epsfig,bm}
\usepackage{graphicx}
\usepackage{amsmath,amssymb,psfrag,textcomp}
\usepackage{amsthm,upgreek}  
\usepackage[geometry]{ifsym}

\newcommand{\be}{\begin{equation}}
\newcommand{\ee}{\end{equation}}
\newcommand{\bd}{\begin{displaymath}}
\newcommand{\ed}{\end{displaymath}}

\newcommand{\erf}{{\rm erf}}

\newcommand{\beeq}[1] {\begin{equation}\begin{split}#1\end{split}\end{equation}}
\newcommand{\beeqn}[1] {\begin{equation*}\begin{split}#1\end{split}\end{equation*}}

\begin{document}
\title{A weighted belief-propagation algorithm to estimate volume-related properties of random polytopes}
\author{Francesc Font-Clos}
\affiliation{Centre de Recerca Matem\`atica, Edifici C, Campus Bellaterra, E-08193 Bellaterra (Barcelona), Spain}
\affiliation{Departament de Matem\`atiques, Universitat Aut\`onoma de Barcelona, Edifici C, E-08193 Bellaterra (Barcelona), Spain}
\author{Francesco Alessandro Massucci}
\affiliation{Departament d'Enginyeria Qu\'\i mica, Universitat Rovira i Virgili, 43007 Tarragona, Spain}
\author{Isaac P\'erez Castillo}
\affiliation{Department of Mathematics, King's College London, WC2R 2LS, United Kingdom}
\begin{abstract}
In this work we introduce a novel weighted message-passing algorithm based on the cavity method to estimate volume-related properties of random polytopes, properties which are relevant in various research fields ranging from  metabolic networks \cite{Kauffman2003}, to neural networks \cite{Gardner1988}, to compressed sensing \cite{Krzakala2011}. Unlike the usual approach consisting in approximating the real-valued cavity marginal distributions by a few parameters, we propose an algorithm to faithfully represent the entire marginal distribution. We explain various alternatives to implement the algorithm and benchmark the theoretical findings by showing concrete applications to random polytopes. The results obtained with our approach are found to be in very good agreement with the estimates produced by the Hit-and-Run algorithm \cite{Smith1984}, known to produce uniform sampling.
\end{abstract}

\maketitle

\section{Introduction}
There are many problems appearing in  various research fields that can be mathematically formalised as having to determine the solution set of a collection of linear equalities or inequalities for real-valued unknown variables. More precisely,  given a rectangular $M\times N$ matrix $\bm{\xi}$ with real entries $\xi^\mu_{i}$, $\mu=1,\ldots,M$ and $i=1,\ldots,N$ and given a vector $\bm{\gamma}\in\mathbb{R}^M$, with entries $\gamma^\mu$ with $\mu=1,\ldots,M$, one looks for the set $V$ of vectors $\bm{x}\in D\subseteq\mathbb{R}^N$ which are solutions to e.g. a set of equalities
\beeq{
\sum_{i=1}^N\xi^\mu_i x_i=\gamma^\mu\,,\quad \mu=1,\ldots,M
\label{eq:equalities}
}
or a set of inequalities
\beeq{
\sum_{i=1}^N\xi^\mu_i x_i\geq\gamma^\mu\,,\quad \mu=1,\ldots,M\,.
\label{eq:inequalities}
}
A first example of a problem related to a set of equalities like \eqref{eq:equalities} is Flux Balance Analysis \cite{Kauffman2003}. This is a technique designed to estimate the reaction rates of a set of chemical reactions in the stationary state. Having in mind future applications in this area, we devote some lines to explain this method. Suppose we have a set of $N$ chemical reactions that produce and consume $M$ chemical substances, and let us denote as $c^\mu$ the concentration of the chemical substance $\mu$. Then we can write down the following mass-balance evolution equations
\begin{equation*}
\frac{d c^\mu(t)}{dt}=\sum_{i=1}^N\xi^\mu_ix_i-\gamma^\mu
\end{equation*}
where $x_i$ is the reaction rate of reaction $i$, $\gamma^\mu$ is an exchange flux of chemical substance $\mu$ with the environment, and  $(\xi^1_i,\ldots,\xi^M_i)$ are the stoichiometric coefficients of reaction $i$. Generally, each reaction rate $x_i$ is a very complicated function of the concentrations and kinetic parameters, i.e.  $x_i=x_i(\bm{c},\text{kinetic coefficients, etc})$. Even in cases in which all these functional relations are known, the resulting equations are highly non-linear and numerically hard to handle. However, if we have good reasons to assume that the system works in the stationary state, that is $\frac{dc^\mu(t)}{dt}=0$, then the reaction rate vector $\bm{x}$ can be considered as an unknown vector to a set of equalities \eqref{eq:equalities}. Besides, due bio-chemical constraints, each reaction rate $x_i$ would we such that $x_i\in[x_{i}^{\text{min}}, x_{i}^{\text{max}}]$ so that $D=\prod_{i=1}^N[x_{i}^{\text{min}},x_{i}^{\text{max}}]$.  Thus, in this context, estimating the volume of solutions corresponds to estimate reaction rates of a set of chemical reactions in the stationary state. Of course, the problem has been transformed from a potentially numerical intractable set of coupled non-linear differential equations, to the more tractable yet very time-consuming task of estimating the volume of solutions $V$ of \eqref{eq:equalities}. Flux Balance Analysis goes one step further and simplifies the problem by replacing the whole volume $V$ by one of its solutions. This is achieved by imposing an objective function which must be optimized (see \cite{Kauffman2003} for details).\\
One example for the case of inequalities \eqref{eq:inequalities} is von Neumann's expanding model for linear economies \cite{VN,DeMartino2005,DeMartino2007}, which, due to its  interesting applications to metabolic networks \cite{DeMartino2009,Martelli2009,DeMartino2010,DeMartino2010b}, deserves a couple of lines of explanation.  In the economic context it was originally introduced, this model assumes that an economy is constituted by $N$ companies consuming and producing $M$ commodities. Each company $i$ is able to operate linearly with a scale of operation $x_i$. If $A=(a_i^\mu)$ and $B=(b_i^\mu)$  are the input and output matrices of such an economy, as some of the input will be used to produce output, the ratio of input to output produced cannot be larger then the a global growth rate $\rho$, that is
\beeq{
\rho\leq\frac{\sum_{i=1}^N b_i^\mu x_i}{\sum_{i=1}^N a_i^\mu x_i}\quad\quad\text{    or    }\quad\quad\sum_{i=1}^N\left( b_i^\mu - \rho a_i^\mu\right) x_i\geq 0\,,\quad\quad \forall \mu=1,\ldots, M\,.
\label{eq:VN}
}
Thus in von Neumann's model of linear economies, one then wonders what are possible values for the vector of operations $\bm{x}$ for a given growth rate, and how can the optimal growth rate be achieved.\\
These two examples are just a glimpse into a myriad of problems, old (e.g. Gardner's optimal capacity problem \cite{Gardner1988}) and new  ones (e.g. compressed sensing \cite{Krzakala2011,Baron2010}), coming from diverse research fields which can be mathematically formalised as either \eqref{eq:equalities} or \eqref{eq:inequalities}, and which have the same goal: find techniques, either analytic or numerical or a combination of both, to estimate the volume of solutions $V$ and statistical properties related to it.\\
Thus, rather than focusing in a research field in particular, in the present paper we analyse the problem on general terms, and, for sake of clarity, we present our new methodology and test it on  simple examples, leaving more complex and certainly more exciting applications, particularly in the area of metabolic networks, for future research.\\
We have organised this paper as follows: in section \ref{sect:MD} we set up the problem at hand and present it in an unifying way. In section \ref{sect:SSM} we apply the cavity method to obtain a set of consistency equations for the cavity marginals. We also discuss a set of consistency equations for the support of these marginals. As we will see, the set of equations for the supports  decouples from the actual shape of the marginals, yielding an extremely simple set of equations, which can be readily solved. Section \ref{sect:WBPA} contains the first main result, consisting in reweighting the cavity equations seeking for an efficient way to solve them. In section \ref{sect:IWBPA} we propose two main ways to implement the weighted cavity equations: the method of the histograms and the weighted population dynamics in the instance. Besides, for the second method we suggest two alternatives: random assignment \&  variable locking  and variable fixing. We have benchmarked all these novel ideas with some examples and reported the results in section \ref{sect:NT}. Further theoretical research and applications, particularly in the area of metabolic networks, are discussed in the conclusions in the last section \ref{sect:con}.
\section{Model Definitions}
\label{sect:MD}
As the analytical treatment to the two problems \eqref{eq:equalities} and \eqref{eq:inequalities} is not that different, we choose to discuss the more general problem related to the set of inequalities \eqref{eq:inequalities}.  We are particularly interested in heterogeneous systems in which the rectangular matrix $\bm{\xi}$ is generally random. This sums up to say quite simply that we focus here in estimating volume-related properties of random polytopes.  In the so-called H-representation, a polytope is defined as the set of points $\bm{x}=(x_1,\ldots,x_N)\in D=D_1\times\cdots\times D_N\subseteq\mathbb{R}^N$  encapsulated by $M$ hyperplanes $\{(\bm{\xi}^\mu,\gamma^\mu)\}_{\mu=1}^M$, with $\bm{\xi}^\mu=(\xi^\mu_1,\ldots,\xi^\mu_N)$ the normal vector of the $\mu$-hyperplane. Its volume can formally be written as
\begin{equation*}
V=\left\{\bm{x}\in D: \sum_{i=1}^N\xi^\mu_i x_i\geq\gamma^\mu\,,\, \mu=1,\ldots,M\right\}\,.
\end{equation*}
From all possible questions related to polytopes in H-representation,  we are particularly interested in that of the volume $V$ and its projection onto each axis, in order words, the single-site marginal pdf $P_i(x_i)$. These two quantities can  mathematically be written as:
\begin{eqnarray}
V = \int_{D} d\bm{x}\prod_{\mu=1}^M \Theta \left[h_\mu(x_{\partial \mu})\right]\label{partition}\,,\quad P_i(x_i)=\frac{1}{V}\int_{D_{\backslash i}} d\bm{x}_{\setminus i}\prod_{\mu=1}^M \Theta \left[h_\mu(x_{\partial \mu})\right]\,,
\end{eqnarray}
where we have defined $h_\mu(x_{\partial \mu})=\sum_{i\in\partial \mu} \xi_i^\mu x_i - \gamma^\mu$, $\Theta(x)$ is the Heaviside step function, and $\bm{x}_{\setminus i}$ denotes the vector $\bm{x}$ without component $x_i$, and $D_{\backslash i}=D_1\times\cdots \times D_{i-1}\times D_{i+1}\times\cdots \times D_{N}$. Notice that, in a strict mathematical sense, the volume of the polytope $V$ defined by eq. \eqref{partition} is always strictly 0, as this is the volume of a $N-M$ dimensional manifold in $N$ dimensions. This  can be formally dealt with as explained in \cite{Braunstein2008}, but as defined here does not impact the results concerning the marginal distributions.
\section{Single-site marginals, cavity equations and supports}
\label{sect:SSM}
Using the cavity method (see appendix \ref{appendixcavity}) it is possible to express the single-site marginals $P_i(x_i)$ in terms of the local variables connected to it. Assuming that the bipartite graph associated to the rectangular matrix $\bm{\xi}$ is locally tree-like, we can arrive to the following set of cavity equations
\begin{eqnarray}
P^{(\nu)}_i(x_i)&=&\frac{1}{V^{(\nu)}_i}\prod_{\mu\in\partial i\setminus\nu}m_\mu^{(i)}(x_i)\label{mtoP}\,,\quad \forall i\,,~ \nu\in\partial i\,,\\
 m_\mu^{(i)}(x_i)&=&\frac{1}{ m_\mu^{(i)}}\int_{D_{\partial \mu\backslash i}} dx_{\partial \mu\backslash i}  \Theta \left(h_\mu(x_{\partial \mu\backslash i})+\xi_i^\mu x_i\right)\prod_{\ell\in\partial \mu\backslash i}P^{(\mu)}_{\ell}(x_\ell)\,,~ \forall i\,,~ \mu\in\partial i\label{Ptom}\,.
\end{eqnarray}
Here the $V^{(\nu)}_i$ and $ m_\mu^{(i)}$ are normalising constants and we have labelled the variable-nodes with Latin indices $i,j,\ldots$ and the factor-nodes with Greek ones $\mu,\nu,\ldots$. Besides, we have used the following standard notation: $\partial i$ denotes the set of factor-nodes neighbouring $i$, $\partial \mu$ denotes the set of variable-nodes neighbouring the factor-node $\mu$, and $x_{\partial \mu}$ is the set of dynamical variables on the set variable-nodes $\partial \mu$. Finally, if $A$ is a set of indices and $i\in A$ then $A\backslash i$ is the set $A$ without index $i$.\\
Once the set of equations \eqref{mtoP} and \eqref{Ptom} is solved, the actual single-site marginals are given in terms of the marginals $\{m_\mu^{(i)}(x_i)\}$:
\begin{eqnarray}
P_i(x_i)&=&\frac{1}{V_i}\prod_{\mu\in\partial i}m_\mu^{(i)}(x_i)\,,
\label{realP}
\end{eqnarray}
with $V_i$ a normalising constant. It is important to point out that the support of the marginal $P_i(x_i)$ is not the  integration domain for  $x_i\in D_i$, but rather the result of intersecting the polytope and $D$, being the intersection then projected onto the $x_i$-axis. Looking at the eqs. \eqref{mtoP} and \eqref{Ptom}, one notices that it is possible to write down self-consistency equations for the support of the marginals in the cavity equations. To do so, let us denote as $R^{(i)}_{\mu}$ and $K^{(\nu)}_{i}$ the support of the marginals $m_\mu^{(i)}(x_i)$ and $P^{(\nu)}_i(x_i)$, respectively. Then from eq. \eqref{mtoP} we can write that
\beeq{
K^{(\nu)}_{i}=\bigcap_{\mu\in\partial i\setminus\nu}R^{(i)}_{\mu}\,.
\label{domain1}
} 
To be able to write down the support $R^{(i)}_{\mu}$ in terms of $K^{(\nu)}_{i}$, we notice from eq. \eqref{Ptom} that, geometrically, we are integrating on a parallelotope $S_{\mu}^{(i)}\subseteq D_{\partial \mu\backslash i}$ defined as the cartesian product of supports $\{K_\ell^{(\mu)}\}_{\ell\in\partial \mu\backslash i}$, that is $S_{\mu}^{(i)}=\times_{\ell\in\partial \mu\backslash i}K_\ell^{(\mu)}$. This parallelotope has $2^{|\partial \mu\backslash i|}$ vertices, whose set we denote as $\mathcal{V}_{\mu}^{(i)}$. As geometrically suggested by eq. \eqref{Ptom}, when $x_i$ is varied the hyperplane $h_\mu(x_{\partial \mu\backslash i})+\xi_i^\mu x_i=0$ will intersect with all these vertices. Let us denote as $\mathcal{I}\left(\mathcal{V}_{\mu}^{(i)}\right)$ the set of $2^{|\partial \mu\backslash i|}$ values of $x_i$ at which this intersection occurs. Then the support $R^{(i)}_{\mu}$ is the maximal set such that
\beeq{
R^{(i)}_{\mu}=\left[\min_{x\in \mathcal{I}} \mathcal{I}\left(\mathcal{V}_{\mu}^{(i)}\right),\max_{x \in \mathcal{I}} \mathcal{I}\left(\mathcal{V}_{\mu}^{(i)}\right)\right]\bigcap R^{(i)}_{\mu}\,.
\label{domain2}
}
Notice that once the collection of supports $R^{(i)}_{\mu}$ are known, then supports $K_i$ for the single-site marginals $P_i(x_i)$ are given by
\beeq{
K_{i}=\bigcap_{\mu\in\partial i}R^{(i)}_{\mu}\,.
\label{domain3}
}
The set of equations \eqref{domain1} and \eqref{domain2} is decoupled from the set of the cavity equations \eqref{mtoP} and \eqref{Ptom}, in the sense that the actual shape of the marginals is not needed to obtain information solely about their support. This rather simple observation is quite relevant for two reasons. Firstly, to apply the algorithms we devise in Sect. \ref{sect:WBPA}  we need to know the support of the marginals beforehand in order to avoid rejection. This will become clearer in the explanation of the method, but in anticipation and with a modest amount of foresight we note that  by simply inspecting  eq. \eqref{domain1}: either $x_i$ lies within the intersection of all the supports $\{R^{(i)}_{\mu}\}_{\mu\in\partial i\setminus\nu}$ or -at least- one of the $m$-functions is zero for that value, and therefore the value $x_i$ should not be allowed. Secondly, as we will see in the derivation below, the self-consistency equations for the supports are much simpler than the corresponding equations for the marginals, meaning that they can be solved extremely fast. Thus, as the supports give valuable information of allowed values for the dynamical variables, these self-consistency equations are important in their own right, as they can be applied, for instance, to study the behaviour of metabolic networks under perturbations (e.g. effect of gene knockout in reaction rates) or, as we illustrate below in section \ref{examplecritical}, to determine the critical line for the global growth rate in Von Neumann's model described in eq. \eqref{eq:VN}.
\subsection{Expressing the self-consistency equations for the supports in terms of endpoints}
We now move on to implement the set of equations \eqref{domain1} and \eqref{domain2} to obtain each support $K_{i}^{(\mu)}$. As the marginals are the result of projecting a convex polytope, we expect them to be singly-supported. We then define $K_{i}^{(\mu)}=[k_{i,-}^{(\mu)},k_{i,+}^{(\mu)}]$ and similarly $R_{\mu}^{(i)}=[r_{\mu,-}^{(i)},r_{\mu,+}^{(i)}]$. From the Heaviside funcion in eq. \eqref{Ptom} we know that $h_\mu(x_{\partial \mu\backslash i})+\xi_i^\mu x_i\geq 0$. This implies that for an auxiliary variable $z_{\mu}$ we can write $h_\mu(x_{\partial \mu\backslash i})+\xi_i^\mu x_i=z_\mu$ or $x_i=(1/\xi_i^\mu )[z_\mu-h_\mu(x_{\partial \mu\backslash i})]$. If the reader finds the introduction of the variable $z_\mu$ rather obscure, an alternative interpretation will be given in Sect. \ref{sect:WBPA}. Next, let us now denote as $x^{i}_{\mu,\sigma_{\partial \mu\backslash i}}$ the $2^{|\partial \mu\backslash i|}$ values of $x_i$ at  each of the vertices of the parallelotope $S_{\mu}^{(i)}=\times_{\ell\in\partial \mu\backslash i}K_\ell^{(\mu)}$, that is:
\beeqn{
x^{(i)}_{\mu,\sigma_{\partial \mu\backslash i}}=\frac{1}{\xi_i^\mu} y^{(i)}_{\mu,\sigma_{\partial \mu\backslash i}}\,,\quad  y^{(i)}_{\mu,\sigma_{\partial \mu\backslash i}}=z_{\mu}+\gamma^\mu-\sum_{\ell\in\partial \mu\backslash i}\xi_\ell^\mu k^{(\mu)}_{\ell,\sigma_{\ell}}
}
To look for the maximum and minimum in the set of values $\{x^{(i)}_{\mu,\sigma_{\partial \mu\backslash i}}\}$, we notice that we do not need to check all $2^{|\partial \mu\backslash i|}$ values, but rather we can write down directly an expression for them. Indeed, let us define first
\beeq{
t^{(i)}_{\mu,-}=z_{\text{min}}+\gamma^\mu-\sum_{\ell\in\partial \mu\backslash i}\max\{\xi_\ell^\mu k^{(\mu)}_{\ell,\sigma_{\ell}}\}_{\sigma_{\ell}\in\{-,+\}}\,,\quad t^{(i)}_{\mu,+}=z_{\text{max}}+\gamma^\mu-\sum_{\ell\in\partial \mu\backslash i}\min\{\xi_\ell^\mu k^{(\mu)}_{\ell,\sigma_{\ell}}\}_{\sigma_{\ell}\in\{-,+\}}
\label{endpoint1}
}
where $z_{\text{min}}=0$ and $z_{\text{max}}$ are the  minimum and maximum values the variable $z_\mu$ can take \footnote{Since the polytope is at most restricted by $D$, then $z_{\mu}$ has indeed a finite maximum value. For practical purposes one simply puts a cutoff maximum value.}. Then the maximum and minimum values of $x_i$ are
\beeq{
o^{(i)}_{\mu,+}=\max\{t^{(i)}_{\mu,\sigma}/\xi_i^\mu\}_{\sigma\in\{-,+\}}\,,\quad \quad o^{(i)}_{\mu,-}=\min\{t^{(i)}_{\mu,\sigma}/\xi_i^\mu\}_{\sigma\in\{-,+\}}
\label{endpoint2}
}
From here we can have finally write
\beeq{
r_{\mu,-}^{(i)}=\max \{o^{(i)}_{\mu,-},r_{\mu,-}^{(i)}\}\,,\quad\quad  r_{\mu,+}^{(i)}=\min \{o^{(i)}_{\mu,+},r_{\mu,+}^{(i)}\}\\
k_{i,-}^{(\nu)}=\max\{r_{\mu,-}^{(i)}\}_{\mu\in\partial i\backslash \nu}\,,\quad\quad k_{i,+}^{(\nu)}=\min\{r_{\mu,+}^{(i)}\}_{\mu\in\partial i\backslash \nu}
\label{endpoint3}
}
The set of equations \eqref{endpoint1}, \eqref{endpoint2}, and \eqref{endpoint3}, is a set of self-consistency cavity equations for the endpoints defining the supports of the cavity marginals. These cavity equations can be solved in the standard manner by using belief-propagation, i.e. fixed-point iteration method, using as a starting support for $K^{(\mu)}_i=D_i=[m_{i,-},m_{i,+}]$ for all $i=1,\ldots,N$ and for all $\mu\in\partial i$. From  eq. \eqref{domain3} and once  the values of $\{r_{\mu,-}^{(i)}\}$ have been estimated, we can calculate the support $K_i$ of the single-site marginals $P_i(x_i)$, \textit{viz.}
\beeqn{
k_{i,-}=\max\{r_{\mu,-}^{(i)}\}_{\mu\in\partial i}\,,\quad\quad k_{i,+}=\min\{r_{\mu,+}^{(i)}\}_{\mu\in\partial i}
}

\section{A weighted belief propagation algorithm}
\label{sect:WBPA}
Before presenting our method let us discuss briefly the various alternatives found in the literature of how to estimate volume-related properties of polytopes. They can be grouped into two main categories: (i) Monte Carlo simulations and (ii) theoretical approaches.\\
By Monte Carlo simulations we generally mean a numerical way of directly sampling  -hopefully uniformly- the volume of a polytope. The Hit-and-Run algorithm \cite{Smith1984}, explained in the appendix \ref{app:har}, is a Monte Carlo method that samples uniformly the volume of a polytope. This method is reasonably efficient only for small dimensions as its mixing time goes as $\mathcal{O}(N^3)$. The MinOver$^{+}$ algorithm \cite{Krauth1987} was originally introduced in neural networks to check whether a solution to a problem of the type \eqref{eq:inequalities} exists. This algorithm was subsequently applied to sample the volume of polytopes in \cite{DeMartino2007,DeMartino2010,Martelli2009,DeMartino2010b}. Unlike the Hit-and-Run algorithm, MinOver$^{+}$ can deal with larger systems, yielding samples that become more uniform the larger the system is.\\ 
By theoretical approaches what we mean is: one identifies the quantities of interest and under not unreasonable assumptions one then tries to write down closed equations for these quantities. This is what we have done here thus far (and also in \cite{DeMartino2007}) by taking as quantities of interests the single-site marginals, finding the set of closed equations \eqref{mtoP} and \eqref{Ptom} by assuming the Bethe approximation for the problem \eqref{eq:inequalities}. For FBA, that is for problems of the type \eqref{eq:equalities} this was done in e.g. \cite{Braunstein2008}, while for the context of compressed sensing similar equations were derived in \cite{Baron2010,Krzakala2011}. Of course the problem is then how to solve the resulting equations (e.g. set of equations \eqref{mtoP} and \eqref{Ptom} in our case) in an efficient way.\\
The reader should note that, in the simplest possible analysis, we expect the computational time associated to solving the cavity equations to go linear with the system size $N$ (as opposed to $\sim N^3$ for the Hit-and-Run algorithm), as this is precisely how the number of cavity equations grows with $N$. Leaving this aside there are, however, more urgent and pressing matters. Indeed, on general terms it is considered a daunting task trying to solve the exact cavity equations by any numerical means when the dynamical variables are continuous, as it is in our case, in the cases \cite{Braunstein2008,Baron2010,Krzakala2011} or, more generally, for any continued-valued spin models on diluted graphs. To overcome this numerical burden, one first approximates the cavity equations somehow and then solves numerically the approximated equations. This can be presented in various ways, but the general approach is first noticing that the marginals can be parametrised by an infinite number of parameters, that is, $P^{(\nu)}_i\left(x_i\Big|\bm{\alpha}^{(\nu)}_i\right)$ and $ m_\mu^{(i)}\left(x_i\Big|\bm{\beta}_{\mu}^{(i)}\right)$ where we have denoted $\bm{\alpha}^{(\nu)}_i=\left(\alpha^{(\nu)}_{i,1},\alpha^{(\nu)}_{i,2},\ldots\right)$ and $\bm{\beta}_{\mu}^{(i)}=\left(\beta_{\mu,1}^{(i)},\beta_{\mu,2}^{(i)},\ldots\right)$. For instance, in this framework the Gaussian approximation means choosing the parameters of the marginals to be their cumulants and assume that only the first two cumulants are different from zero.\\
However, for this type of problems consisting in counting the number of solutions of a set of either linear equalities or inequalites, as it is our case, it is possible to tackle the cavity equations \eqref{mtoP} and \eqref{Ptom} without approximation. To whet the appetite for the forthcoming discussion let us first rewrite the eq. \eqref{Ptom} as follows:
\begin{eqnarray}
 m_\mu^{(i)}(x_i)&=&\frac{1}{ m_\mu^{(i)}}\int_{\mathbb{R}^{+}} dz_\mu\int_{D_{\partial \mu\backslash i}} dx_{\partial \mu\backslash i}  \delta \left(h_\mu(x_{\partial \mu\backslash i})+\xi_i^\mu x_i-z_\mu\right)\prod_{\ell\in\partial \mu\backslash i}P^{(\mu)}_{\ell}(x_\ell)\,,~ \forall i\,,~ \mu\in\partial i\label{Ptom2}
\end{eqnarray}
where we have expressed the Theta functions in terms of a Dirac delta using the integral identity $\theta(x-a)=\int_{\mathbb{R}^{+}} dy\delta[(x-a)-y]$. Note that the new integration variable $z_\mu$ can be understood as a random variable with uniform density in a subset of $\mathbb{R}^{+}$. Indeed, due to the constraints of the problem at hand, the random variable $z_\mu$ will not generally take values on the whole positive line but rather on a compact set of it, and it will be therefore normalisable.\\
Looking at the cavity equation \eqref{Ptom2} we notice that, written in this way, the Dirac delta suggests to estimate the integral expression by using the method of population dynamics (Nb. population dynamics is a smart way to estimate the integral using a Monte Carlo method when the integrand is unknown) as in, for instance, the ensemble cavity equations for discrete spin models on locally treelike graphs. There are two important differences, though: (i) the cavity equation \eqref{Ptom2} is still on the instance; (ii) not all updates $x_i\leftarrow(z_\mu-(h_\mu(x_{\partial \mu\backslash i}))/\xi_i^\mu$ suggested by eq. \eqref{Ptom2} are allowed as there may be integration regions such that $x_i\not\in R^{(i)}_{\mu}$ and therefore the update must be rejected. The first difference is unimportant, the second one is not: if we do not find a way to avoid the rejection region, the resulting population dynamics performed on the instance will be quite inefficient.\\
Fortunately, there is a way to avoid the rejection region completely. To illustrate how, we set aside all notational complicacies of the cavity equations and focus on the problem at hand with the following simpler example:
\beeqn{
m(x)=\frac{1}{m}\int_{[0,1]^n}\left[\prod_{i=1}^n dy_i \rho_i(y_i)\right]\delta \left(x+\sum_{i=1}^n a_iy_i-\gamma\right)\,,\quad x\geq 0
}
where $\rho_i(y_i)$ for $i=1,\ldots,n$ are some arbitrary pdf with $y_i\in [0,1]$.  We note that this integral is zero unless $x=\gamma-\sum_{i=1}^n a_iy_i\geq 0$. Thus the effective integration region is the one enclosed by the hypercube $[0,1]^n$ and the hyper-plane $\gamma=\sum_{i=1}^n a_iy_i$. Deciding to do the integral in a specific order we write
\beeqn{
m(x)=\frac{1}{m}\int_{\mathcal{R}_1(\gamma)} dy_1\rho_1(y_1)\int_{\mathcal{R}_2(y_1,\gamma)} dy_2\rho_2(y_2)\cdots\int_{\mathcal{R}_n(y_1,\ldots, y_{n-1},\gamma)} dy_n\rho_n(y_n)\delta \left(x+\sum_{i=1}^n a_iy_i-\gamma\right)
}
As we want to evaluate the above integral by Monte Carlo, we need to reweight each density $\rho_i$ on the new region $\mathcal{R}_i(y_1,\ldots, y_{i-1},\gamma)$. We write
\beeqn{
\rho_i(y_i|y_1,\ldots, y_{i-1},\gamma)=\frac{\rho_i(y_i)}{w_i(y_1,\ldots, y_{i-1},\gamma)}\,,\quad w_i(y_1,\ldots, y_{i-1},\gamma)& \equiv \int_{\mathcal{R}_i(y_1,\ldots, y_{i-1},\gamma)}d y_i\rho_i(y_i)
}
Thus the expression for $m(x)$ becomes:
\beeqn{
m(x)=\frac{1}{m}\left[\prod_{i=1}^n\int_{\mathcal{R}_i(y_1,\ldots,y_{i-1},\gamma)} dy_i\rho_i(y_i|y_1,\ldots, y_{i-1},\gamma)\right]\left[\prod_{i=1}^nw_i(y_1,\ldots, y_{i-1},\gamma)\right]\delta \left(x+\sum_{i=1}^n a_iy_i-\gamma\right)
}
In this new form the integral  $m(x)$ can be evaluated by Monte Carlo without rejection. This is done as follows:
\begin{enumerate}
\item draw $y_1$ according to $\rho_1(y_1|\gamma)$, draw $y_2$ according to $\rho_2(y_2|y_1,\gamma)$, etc, and finally draw $y_n$ according to $\rho_n(y_n|y_1,\ldots, y_{n-1}, \gamma)$
\item Assign $x\leftarrow \gamma-\sum_{i=1}^n a_iy_i$ with weight $\Omega\equiv\prod_{i=1}^n w_i(y_1,\ldots, y_{i-1},\gamma)$
\end{enumerate}
This process can be then repeated $\mathcal{N}_{p}$ times to obtain a collection of pairs $\{(x_\alpha,\Omega_\alpha)\}_{\alpha=1,\ldots,\mathcal{N}_{p}}$, from which we can reconstruct $m(x)$ as $m(x)\approx\frac{1}{m}\sum_{\alpha=1}^{\mathcal{N}_p}\Omega_\alpha\delta(x-x_\alpha)$  with $m=\sum_{\alpha=1}^{\mathcal{N}_p}\Omega_\alpha$. This reweighting technique avoids the rejection region allowing us to estimate the involved integrals much more efficiently.\\
Thus, if we apply this reweighting technique to cavity equations \eqref{Ptom2},  we obtain
\beeq{
m_\mu^{(i)}(x_i) &=\frac{1}{m_\mu^{(i)}} \left[\prod_{j=1}^{k_\mu^{(i)}}\int_{\mathcal{R}^{(\mu)}_j(x_{\ell_1},\ldots,x_{\ell_{j-1}},\gamma^\mu)}  d x_{\ell_{j}} P^{(\mu)}_{\ell_j}(x_{\ell_j}|x_{\ell_1},\ldots,x_{\ell_{j-1}},\gamma^\mu)\right]\int_{\mathcal{R}^{(\mu)}_j(x_{\partial \mu\backslash i},\gamma^\mu)} \frac{dz_\mu}{\omega^{(\mu)}(x_{\partial \mu\backslash i},\gamma^\mu)}\\
&\times ~\delta\left(h_\mu(x_{\partial \mu\backslash i})+\xi_i^\mu x_i-z_\mu\right) \left[\prod_{j=1}^{k_\mu^{(i)}}\omega_{j}^{(\mu)}(x_{\ell_1},\ldots,x_{\ell_{j-1}},\gamma^\mu)\right]\omega^{(\mu)}(x_{\partial \mu\backslash i},\gamma^\mu)\,,~ \forall i\,,~ \mu\in\partial i
\label{PtomReW}
}
where we have used the notation  $k_\mu^{(i)}=|\partial \mu \setminus i|$ and $\ell_{j}\in \partial \mu \setminus i$ for $j=1,\ldots,k_\mu^{(i)}$ and we have defined
\begin{eqnarray*}
P^{(\mu)}_{\ell_j}(x_{\ell_j}|x_{\ell_1},\ldots,x_{\ell_{j-1}},\gamma^\mu)&=&\frac{P^{(\mu)}_{\ell_j}(x_{\ell_j})}{\omega_{j}^{(\mu)}(x_{\ell_1},\ldots,x_{\ell_{j-1}},\gamma^\mu)}\,,\\
 \omega_{j}^{(\mu)}(x_{\ell_1},\ldots,x_{\ell_{j-1}},\gamma^\mu)&=&\int_{\mathcal{R}^{(\mu)}_j(x_{\ell_1},\ldots,x_{\ell_{j-1}},\gamma^\mu)}dx_{\ell_j}P^{(\mu)}_{\ell_j}(x_{\ell_j})\\
\omega^{(\mu)}(x_{\partial \mu\backslash i},\gamma^\mu)&=&\int_{\mathcal{R}^{(\mu)}_j(x_{\partial \mu\backslash i},\gamma^\mu)} dz_\mu= \left|\mathcal{R}^{(\mu)}_j(x_{\partial \mu\backslash i},\gamma^\mu)\right|
\end{eqnarray*}
Although we could have chosen any order of integration to write down eq. \eqref{PtomReW}, it is convenient, as we will explain below, to integrate the variable $z_\mu$ the first one \footnote{Note that while $z_\mu$ is the variable to formally be integrated the first one, within our Monte Carlo method, $z_\mu$ is the last variable to be estimated.}. We will henceforth generally call the set of equations \eqref{PtomReW} and \eqref{mtoP} as  \textit{the weighted cavity equations} and the algorithm to solve them  \textit{weighted belief-propagation} or \textit{weighted message-passing algorithm}.
\section{Implementing the weighted belief-propagation algorithm} 
\label{sect:IWBPA}
We move on to discuss the actual implementation of the algorithm to numerically solve the set of equations \eqref{mtoP} and \eqref{PtomReW}. As we will see shortly, although we have found a nice method to avoid rejection, further complicacies appear in the horizon which may put into question the discussion we have had thus far. Fortunately, we describe a way to overcome them.\\
We present here two alternative implementations: \textit{the method of histograms} and \textit{the weighted population dynamics in the instance}.
\subsection{Method of histograms}
Ideally, to solve numerically the weighted cavity equations we would firstly represent the marginals $P_i^{(\mu)}(x_i)$ and $m_{\mu}^{(i)}(x_i)$ with populations of pairs $\left\{\left(w_{i,\alpha}^{(\mu)},x_{i,\alpha}^{(\mu)}\right)\right\}_{\alpha=1,\ldots,\mathcal{N}_p}$ and  $\left\{\left(v_{i,\alpha}^{(\mu)},x_{\mu,\alpha}^{(i)}\right)\right\}_{\alpha=1,\ldots,\mathcal{N}_p}$, respectively, so that $P_i^{(\mu)}(x_i)\approx(1/P_i^{(\mu)})\sum_{\alpha=1}^{\mathcal{N}_p}w_{i,\alpha}^{(\mu)}\delta\left(x_i-x_{i,\alpha}^{(\mu)}\right) $ and $m_\mu^{(i)}(x_i)\approx(1/m_\mu^{(i)})\sum_{\alpha=1}^{\mathcal{N}_p}v_{\mu,\alpha}^{(i)}\delta\left(x_i-x_{\mu,\alpha}^{(i)}\right)$, where $P_i^{(\mu)}$ and $m_\mu^{(i)}$ are normalisation factors. There is a drawback though: while it is possible -and desirable- to use the population of the $P$'s to evaluate eq. \eqref{PtomReW}  updating the population for the $m$'s,  it is difficult to see how to use directly this representation to update the population for the $m$'s according to eq. \eqref{mtoP}. Of course, one could always construct histograms for the set of functions $\{m_\mu^{(i)}(x_i)\}_{\mu\in\partial i/\nu}$ to calculate a histogram for $P^{(\nu)}_i(x_i)$ according to eq. \eqref{mtoP}, and then draw a population for $P^{(\nu)}_i(x_i)$ from its histogram. But if we need to construct histograms at some point in the algorithm, it is actually a waste of time to go back and forth between a representation of histograms  and a representation of populations.\\
Thus, due to equation \eqref{mtoP}, we are unfortunately obliged to implement a weighted message-passing algorithm where the functions $P_i^{(\mu)}(x_i)$ and $m_{\mu}^{(i)}(x_i)$ are quite simply represented by histograms. We call this implementation \textit{the method of histograms}.\\
Discretisation of the unweighted cavity equations and their Fourier transform was the method used in \cite{Braunstein2008,Baron2010} in the context of FBA and compressive sensing. Let us see how it is possible to avoid discretisation altogether.
\subsection{Weighted population dynamics in the instance}
Fortunately, it is possible to overcome the previous problem of having to go through the $m$-functions via equation \eqref{mtoP}. To illustrate the method we again ignore all tediousness regarding notation and analyse the following simpler example that captures the complicacies arising from combining equations \eqref{mtoP} and \eqref{PtomReW}, namely, the appearance of multiple Dirac deltas:
\beeqn{
r(x)=\frac{1}{r}\int_{[0,1]^n}\left[\prod_{i=1}^n dy_i \rho_i(y_i)\right]\delta \left(x+\sum_{i=1}^n a_iy_i-\gamma\right)\int_{[0,1]^m}\left[\prod_{i=1}^m dz_i \psi_i(z_i)\right]\delta \left(x+\sum_{i=1}^m b_iz_i-\delta\right)\,,
}
with $x\geq 0$ and where $\rho_i(y_i)$, $i=1,\ldots,n$ and $\psi_i(z_i)$, $i=1,\ldots,m$ are some arbitrary pdf with $y_i,z_i\in [0,1]$. As we have already discussed the problem of rejection, we first simply reweight this equation accordingly:
\beeq{
r(x)&=\frac{1}{r}\left[\prod_{i=1}^n\int_{\mathcal{R}^{(\rho)}_i(y_1,\ldots,y_{i-1},\gamma)} dy_i\rho_i(y_i|y_1,\ldots, y_{i-1},\gamma)\right]\left[\prod_{i=1}^nw^{(\rho)}_i(y_1,\ldots, y_{i-1},\gamma)\right]\delta \left(x+\sum_{i=1}^n a_iy_i-\gamma\right)\\
&\times\left[\prod_{i=1}^m\int_{\mathcal{R}^{(\psi)}_i(z_1,\ldots,z_{i-1},\gamma)} dz_i\psi_i(z_i|z_1,\ldots, z_{i-1},\delta)\right]\left[\prod_{i=1}^m w^{(\psi)}_i(z_1,\ldots, z_{i-1},\delta)\right]\delta \left(x+\sum_{i=1}^m b_iz_i-\delta\right)
\label{example2}
}
As there are two Dirac deltas in the expression \eqref{example2}, it is natural to ask how resolve the issue of assignment for the variable $x$ so that we can estimate the integrals by Monte Carlo. We describe two options: \textit{random assignment \&  variable locking} and \textit{variable fixing}.
\subsubsection{Random Assignment \&  Variable Locking}
In this case we use one of the Dirac deltas to assign a value to $x$, locking such value in the other Dirac deltas. Suppose, for instance, that in our example we use the first Dirac delta to assign a value to $x$. The first steps of the algorithm would be almost as before:
\begin{enumerate}
\item draw $y_1$ according to $\rho_1(y_1|\gamma)$, draw $y_2$ according to $\rho_2(y_2|y_1,\gamma)$, etc, and finally draw $y_n$ according to $\rho_n(y_n|y_1,\ldots, y_{n-1}, \gamma)$
\item Assign $x\leftarrow x^{\star}=\gamma-\sum_{i=1}^n a_iy_i$
\item Calculate the weight $\Omega\equiv\prod_{i=1}^n w^{(\rho)}_i(y_1,\ldots, y_{i-1},\gamma)$
\end{enumerate}
In the second Dirac delta, the value of $x$ has been locked to $x^\star$. The integration regions will depend on $x^\star$ and we denote this by writing $\mathcal{R}^{(\psi)}_i(z_1,\ldots,z_{i-1},\gamma,x^\star)$. The multiple integral over $\bm{z}$ is then done as follows:
\begin{enumerate}
\item draw $z_1$ according to $\psi_1(z_1|\delta,x^\star)$, draw $z_2$ according to $\psi_2(z_2|z_1,\delta,x^\star)$, etc, and finally draw $z_{m-1}$ according to $\psi_{m-1}(z_{m-1}|z_1,\ldots, z_{m-2}, \delta,x^\star)$.
\item The Dirac delta locks the value $z_m\leftarrow(\delta-x^\star-\sum_{i=1}^{m-1} b_iz_i)/b_m$ allowing us to integrate the last integral over $z_m$.
\item Calculate the weight $\Gamma\equiv\psi_{m}(z_m|z_1,\ldots,z_{m-1},\delta,x^\star)\prod_{i=1}^mw^{(\psi)}_i(z_1,\ldots, z_{i-1},\delta,x^\star)=\psi_{m}(z_m)\prod_{i=1}^{m-1}w^{(\psi)}_i(z_1,\ldots, z_{i-1},\delta,x^\star)$
\item Assign $x^\star$ an overall weight $\Delta\equiv\Omega \Gamma$
\end{enumerate}
As before this process can be repeated $\mathcal{N}_{p}$ times to obtain a collection of pairs $\{(x^\star_\alpha,\Delta_\alpha)\}_{\alpha=1,\ldots,\mathcal{N}_{p}}$, from which we can reconstruct $r(x)$ as $r(x)\approx\frac{1}{r}\sum_{\alpha=1}^{\mathcal{N}_p}\Delta_\alpha\delta(x-x^\star_\alpha)$ with $r=\sum_{\alpha=1}^{\mathcal{N}_p}\Delta_\alpha$.

\subsubsection{Variable fixing}
In this case we fix the value of $x$ from the beginning, then evaluate each integral according to the second part of the algorithm before. For our example at hand the steps are:
\begin{enumerate}
\item Fix a value of $x$ to $x^\star$ within its support for $r(x)$.
\item For the integrals involving function $\rho$:
\begin{enumerate}
\item draw $y_1$ according to $\rho_1(y_1|\gamma,x^\star)$, draw $y_2$ according to $\rho_2(y_2|y_1,\gamma,x^\star)$, etc, and finally draw $y_{n-1}$ according to $\rho_{n-1}(y_{n-1}|y_1,\ldots, y_{n-2}, \gamma,x^\star)$.
\item The Dirac delta locks the value $y_n\leftarrow(\delta-x-\sum_{i=1}^{n-1} a_iz_i)/a_n$ allowing us to integrate the last integral over $z_n$.
\item Calculate the weight $\Gamma_{\rho}\equiv\psi_{n}(y_n|y_1,\ldots,y_{n-1},\gamma,x^\star)\prod_{i=1}^nw^{(\rho)}_i(y_1,\ldots, y_{i-1},\gamma,x^\star)=\rho_{n}(y_n)\prod_{i=1}^{n-1}w^{(\rho)}_i(y_1,\ldots, y_{i-1},\gamma,x^\star)$
\end{enumerate}
\item Repeat same process for function $\psi$ and calculate the weight $\Gamma_{\psi}$
\item Assign a weight $\Delta\equiv\Gamma_{\rho}\Gamma_{\psi}$ to the fixed value $x^\star$
\end{enumerate}
This process can be repeated $\mathcal{N}_p$ times, scanning the support of $r(x)$ to obtain a collection of pairs $\{(x_\alpha,\Delta_\alpha)\}_{\alpha=1,\ldots,\mathcal{N}_{p}}$, from which we can reconstruct $r(x)$ as $r(x)\approx\frac{1}{r}\sum_{\alpha=1}^{\mathcal{N}_p}\Delta_\alpha\delta(x-x_\alpha)$ with  $r=\sum_{\alpha=1}^{\mathcal{N}_p}\Delta_\alpha$.\\
This implementation has two clear advantages with respect the previous one. The first one is that since we are not using the Dirac delta to randomly assign a value of $x$, this can be used to perform one less integral. For this reason, we have found more convenient to arrange the order of integration in eq. \eqref{PtomReW}, so that the integral over $z_\mu$ is the last one to be estimated by the algorithm, so that the contributing weight is the unity. The second advantage lies on the fact that at a fix value of $x$ what we should obtain is an averaged value of the weight. The weight is itself a random variable whose variance, even in simple cases, can be rather large. Thus it is appropriate to refine the estimates of the weights by replacing them by averaged values. To do this we simply calculate $T$ estimates for each weight $\{\Gamma_{\rho,t}\}_{t=1,\ldots,T}$ and $\{\Gamma_{\psi}\}_{t=1,\ldots,T}$, calculate the average weights $\overline{\Gamma_{\rho}}=\frac{1}{T}\sum_{t=1}^T\Gamma_{\rho,t}$ and $\overline{\Gamma_{\psi}}=\frac{1}{T}\sum_{t=1}^T\Gamma_{\psi,t}$ and assign instead the weight $\overline{\Gamma_{\rho}}\,\overline{\Gamma_{\psi}}$ to the fixed value $x^\star$ \footnote{Clearly, $T$ does not have to be fixed, but chosen so as to achieve certain accuracy, nor does it have to be the same for all weights.}.
\subsubsection{Implementation to our problem}
 Considering all the points discussed previously, while hoping that the tedious but necessary notation we are about to use does not distract the reader, the core of the algorithm for the weighted population dynamics to solve the cavity equations \eqref{mtoP} and \eqref{PtomReW} is as follows. We assume that all supports $K_{i}^{(\mu)}$ for each marginal $P_{i}^{(\mu)}(x_i)$ are known and we represent each marginal by a population of pairs $\left\{\left(w_{i,\alpha}^{(\mu)},x_{i,\alpha}^{(\mu)}\right)\right\}_{\alpha=1,\ldots,\mathcal{N}_p}$. Then for the \textit{random assignment \& variable locking} the essential steps are the following:
\begin{enumerate}
\item Choose a  variable-node $i$, a factor-node $\nu\in\partial i$, and a value $x_i\in K_{i}^{(\nu)}$
\item Chose a factor-node  $\mu_{0}\in\partial i/\nu$
\begin{enumerate}
\item Draw $x_{\ell_1}$ with probability $P^{(\mu_0)}_{\ell_1}(x_{\ell_1}|\gamma^{\mu_0})$, draw $x_{\ell_2}$ with probability $P^{(\mu_0)}_{\ell_2}(x_{\ell_2}|x_{\ell_1},\gamma^{\mu_0})$, etc, draw $x_{\ell_{k_{\mu_0}^{(i)}}}$ with probability $P^{(\mu_0)}_{\ell_{k_{\mu_0}^{(i)}}}(x_{\ell_1}|x_{\ell_1},\ldots,x_{\ell_{k_{\mu_0}^{(i)}-1}},\gamma^{\mu_0})$, and draw a uniform random variable $z^\mu$ in the segment $\mathcal{R}^{(\mu_0)}_j(x_{\partial \mu_0\backslash i},\gamma^{\mu_0})$
\item Assign $x_i\leftarrow [z_{\mu_0}-h_\mu(x_{\partial \mu_0\backslash i})]/\xi_i^{\mu_0}$

\item Calculate $\Omega_{\mu_0}^{(i)}=\omega^{(\mu)}(x_{\partial \mu\backslash i},\gamma^\mu)\prod_{j=1}^{k_\mu^{(i)}}\omega_{j}^{(\mu)}(x_{\ell_1},\ldots,x_{\ell_{j-1}},x_i,\gamma^\mu)$
\end{enumerate}
\item For all $\mu\in\partial i/(\nu\cup\mu_0)$
\begin{enumerate}
\item Draw $x_{\ell_1}$ with probability $P^{(\mu)}_{\ell_1}(x_{\ell_1}|x_i,\gamma^\mu)$, draw $x_{\ell_2}$ with probability $P^{(\mu)}_{\ell_2}(x_{\ell_2}|x_{\ell_1},x_i,\gamma^\mu)$, etc, draw $x_{\ell_{k_\mu^{(i)}}}$ with probability $P^{(\mu)}_{\ell_{k_\mu^{(i)}}}(x_{\ell_1}|x_{\ell_1},\ldots,x_{\ell_{k_\mu^{(i)}-1}},x_i,\gamma^\mu)$
\item Calculate $\Omega_{\mu}^{(i)}=\prod_{j=1}^{k_\mu^{(i)}}\omega_{j}^{(\mu)}(x_{\ell_1},\ldots,x_{\ell_{j-1}},x_i,\gamma^\mu)$\end{enumerate}
\item Calculate $\Gamma_{i}^{(\nu)}=\prod_{\mu\in\partial i/\nu}\Omega^{(i)}_{\mu}$ and replace a pair of the population of the marginal $P^{(\nu)}_i(x_i)$ by the new  pair $\left(\Gamma_{i}^{(\nu)},x_i\right)$
\end{enumerate}
\noindent For the second implementation, that is \textit{variable fixing}, we enumerate the following basic steps:
\begin{enumerate}
\item Choose a  variable-node $i$, a factor-node $\nu\in\partial i$, and a value $x_i\in K_{i}^{(\nu)}$
\item For all $\mu\in\partial i/\nu$
\begin{enumerate}
\item For $t=1,\ldots, T$
\begin{enumerate}
\item Draw $x_{\ell_1}$ with probability $P^{(\mu)}_{\ell_1}(x_{\ell_1}|x_i,\gamma^\mu)$, draw $x_{\ell_2}$ with probability $P^{(\mu)}_{\ell_2}(x_{\ell_2}|x_{\ell_1},x_i,\gamma^\mu)$, etc, draw $x_{\ell_{k_\mu^{(i)}}}$ with probability $P^{(\mu)}_{\ell_{k_\mu^{(i)}}}(x_{\ell_1}|x_{\ell_1},\ldots,x_{\ell_{k_\mu^{(i)}-1}},x_i,\gamma^\mu)$
\item Calculate $w_{t}=\prod_{j=1}^{k_\mu^{(i)}}\omega_{j}^{(\mu)}(x_{\ell_1},\ldots,x_{\ell_{j-1}},x_i,\gamma^\mu)$
\end{enumerate}
\item Set $\Omega^{(i)}_{\mu}=\frac{1}{T}\sum_{t=1}^T w_t$
\end{enumerate}
\item Calculate $\Gamma_{i}^{(\nu)}=\prod_{\mu\in\partial i/\nu}\Omega^{(i)}_{\mu}$ and replace a pair of the population of the marginal $P^{(\nu)}_i(x_i)$ by the new  pair $\left(\Gamma_{i}^{(\nu)},x_i\right)$
\end{enumerate}
It is important to notice here that in the last step the replacement must be done according to how the value of $x_i$ is selected on the first place, otherwise we could generate a biased sampling of the support $K_{i}^{(\nu)}$, that is, if in the first step $x_i$ is chosen uniformly randomly, or fixed then the replacement must be done in the same manner. Besides, instead of replacing other alternatives would be, for instance, to increase the populations in those regions where the sampling is rather poor, or to use a non-uniform sampling taking more points wherever they are most needed.

\section{Numerical Tests}
\label{sect:NT}
To illustrate the previously discussed results we apply them to some simple examples.
\subsection{A toy example regarding supports of marginals}
\begin{figure}[h]
\begin{picture}(350,180)
\put(71,61){$x$}
\put(115,115){$y$}
\put(50,115){$z$}
\put(77,10){$\,_{x\geq 1}$}
\put(65,122){$\,_{x+y+z\leq 4}$}
\put(12,160){$\,_{z\geq 1}$}
\put(140,160){$\,_{y\geq 1}$}
\put(0,0){\includegraphics[width=6cm, height=6cm]{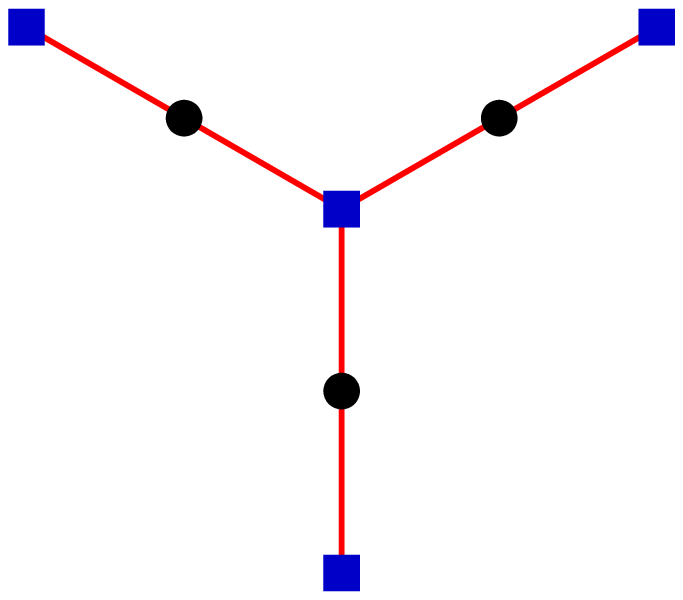}}
\put(200,0){\includegraphics[width=6cm, height=6cm]{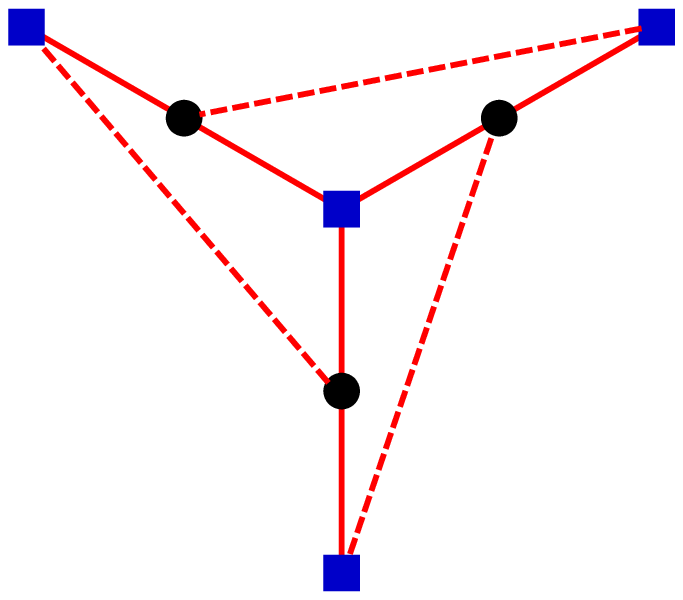}}
\put(271,61){$x$}
\put(315,115){$y$}
\put(250,115){$z$}
\put(262,10){$\,_{x-\epsilon y\geq 1-\epsilon}$}
\put(265,122){$\,_{x+y+z\leq 4}$}
\put(198,160){$\,_{z-\epsilon x\geq 1-\epsilon}$}
\put(327,160){$\,_{y-\epsilon z\geq 1-\epsilon}$}
\end{picture}
\caption{Bipartite graphs associated to the polyhedron defined in our example \eqref{eq:toyexample1}. The left one it is exactly a tree  corresponding for $\epsilon=0$. The right one corresponds to the generic loopy case of $\epsilon\neq0$.}
\label{fig1}
\end{figure}

\begin{figure}[h]
\begin{center}
\includegraphics[width=8cm, height=5cm]{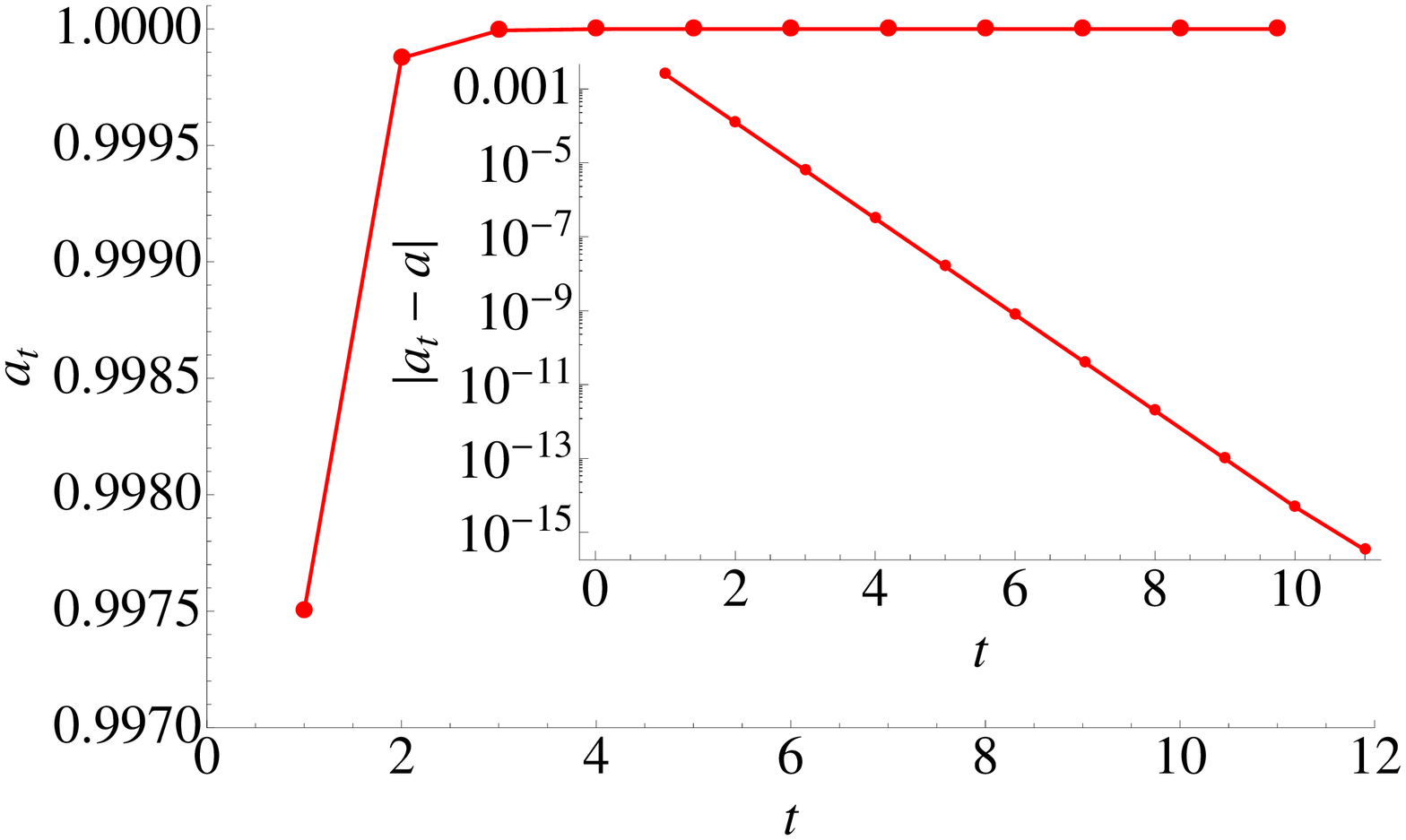}\quad \includegraphics[width=8cm, height=5cm]{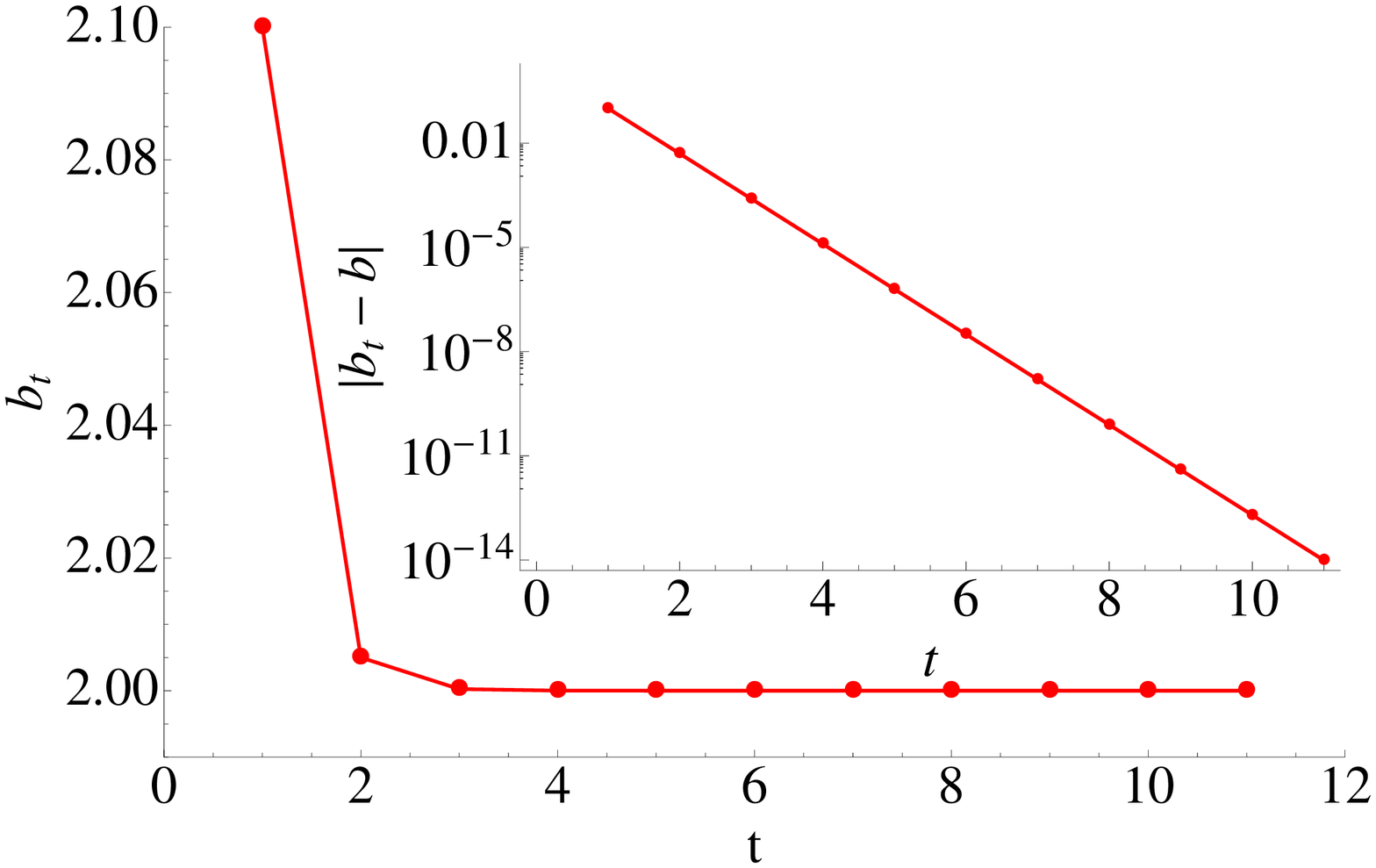}
\end{center}
\caption{Results of iterating the cavity equations \eqref{endpoint1}, \eqref{endpoint2}, and \eqref{endpoint3} for the supports. Due to symmetry of the problem, the support and marginals is the same for the three variables and we parametrise the support as $[a,b]$. We have taken a value of $\epsilon=0.05$. Left (right) figure corresponds to the evolution of the cavity equations for the lower endpoint $a$ ($b$). The inset reports the absolute error between the value of the endpoint at iteration $t$ and the exact value.}
\label{fig2}
\end{figure}

Consider the polyhedron described by the following set of inequalities $\bm{\xi}_\epsilon\bm{x}\geq\bm{\gamma}$ with
\beeq{
\bm{\xi}_{\epsilon}=\begin{pmatrix}
-1&-1&-1\\
1&-\epsilon&0\\
0&1&-\epsilon\\
-\epsilon&0&1\\
\end{pmatrix}\,,\quad \quad \bm{x}=\begin{pmatrix}
x\\
y\\
z
\end{pmatrix}\,,\quad \quad \bm{\gamma}=\begin{pmatrix}
-4\\
1-\epsilon\\
1-\epsilon\\
1-\epsilon
\end{pmatrix}
\label{eq:toyexample1}
}
For $\epsilon=0$ the associated graph to $\bm{\xi}_\epsilon$ is exactly a tree, as shown in figure \ref{fig1}, so that iteration of the cavity equations for the support converge at one step. On the other side, if $\epsilon\neq 0$ the graph is loopy and, depending on the value of $\epsilon$ the estimates from the cavity equations can be very bad indeed.
Yet for small $\epsilon$ one obtains very precise results. In figure \ref{fig2} we plotted the results of iterating the cavity equations \eqref{endpoint1}, \eqref{endpoint2}, and \eqref{endpoint3} for the supports. Due to symmetry of the problem, the support and marginals is the same for the three variables and we parametrise the support as $[a,b]$. We have taken a value of $\epsilon=0.05$.\\
It is important to notice that as the dynamical variables are continous we have at our disposal an ample set of transformations we can perform on the system. This can be used to transform a very loopy network into a more tree-like one. For instance, in this very simple example we notice that we can write $\bm{\xi}_\epsilon=\bm{\eta}_\epsilon\bm{T}_\epsilon$ with 
\beeqn{
\bm{\eta}_{\epsilon}=\begin{pmatrix}
-\frac{1}{1-\epsilon}&-\frac{1}{1-\epsilon}&-\frac{1}{1-\epsilon}\\
1&0&0\\
0&1&0\\
0&0&1\\
\end{pmatrix}\,,\quad  \bm{T}_\epsilon=\begin{pmatrix}
1&-\epsilon&0\\
0&1&-\epsilon\\
-\epsilon&0&1\\
\end{pmatrix}
}
Then, the transformation $\bm{y}=\bm{T}_\epsilon\bm{x}$ takes the system into an exact tree. Of course, under which general conditions it is possible to  perform these type of transformation is not clear to us just yet, nor it is clear which are the most appropriate transformations, whose inverse transformation can be performed efficiently.

\subsection{The critical line $\rho(n)$ in Von Neumann's model}
\begin{figure}[h]
\begin{picture}(500,200)
\put(0,0){\includegraphics[width=9cm, height=7cm]{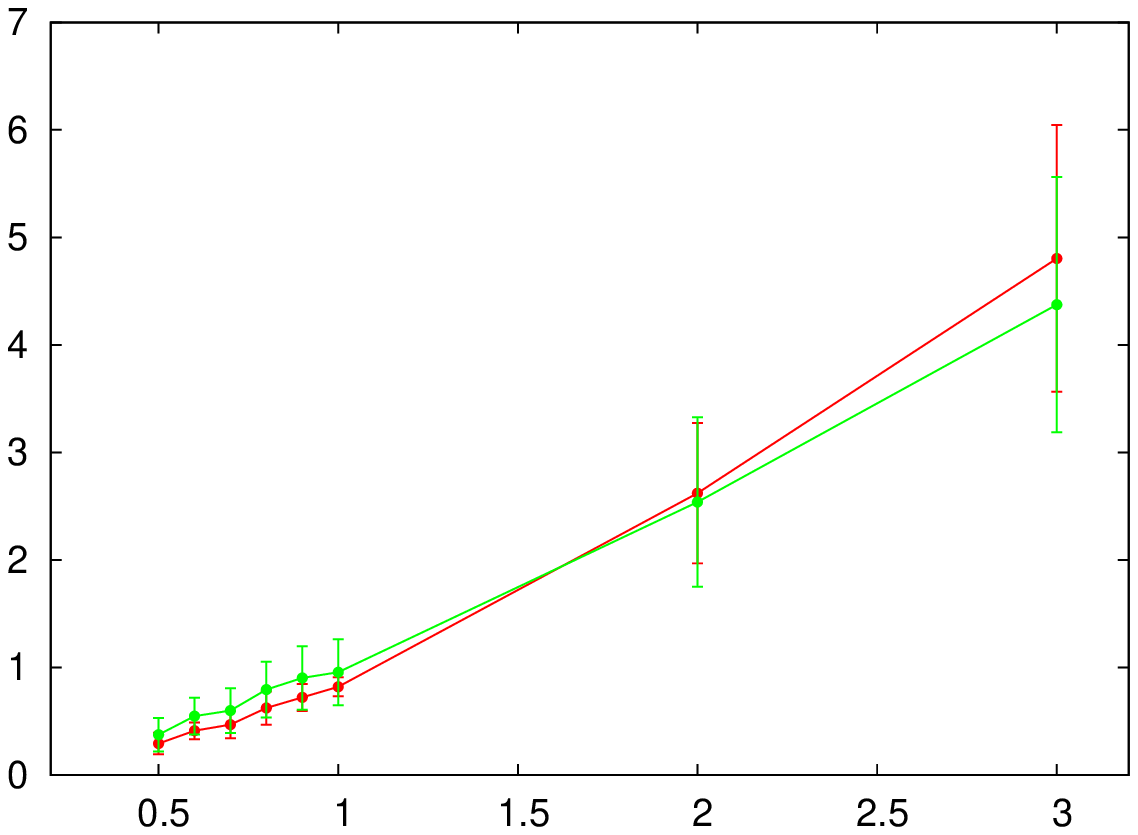}}
\put(250,0){\includegraphics[width=9cm, height=7cm]{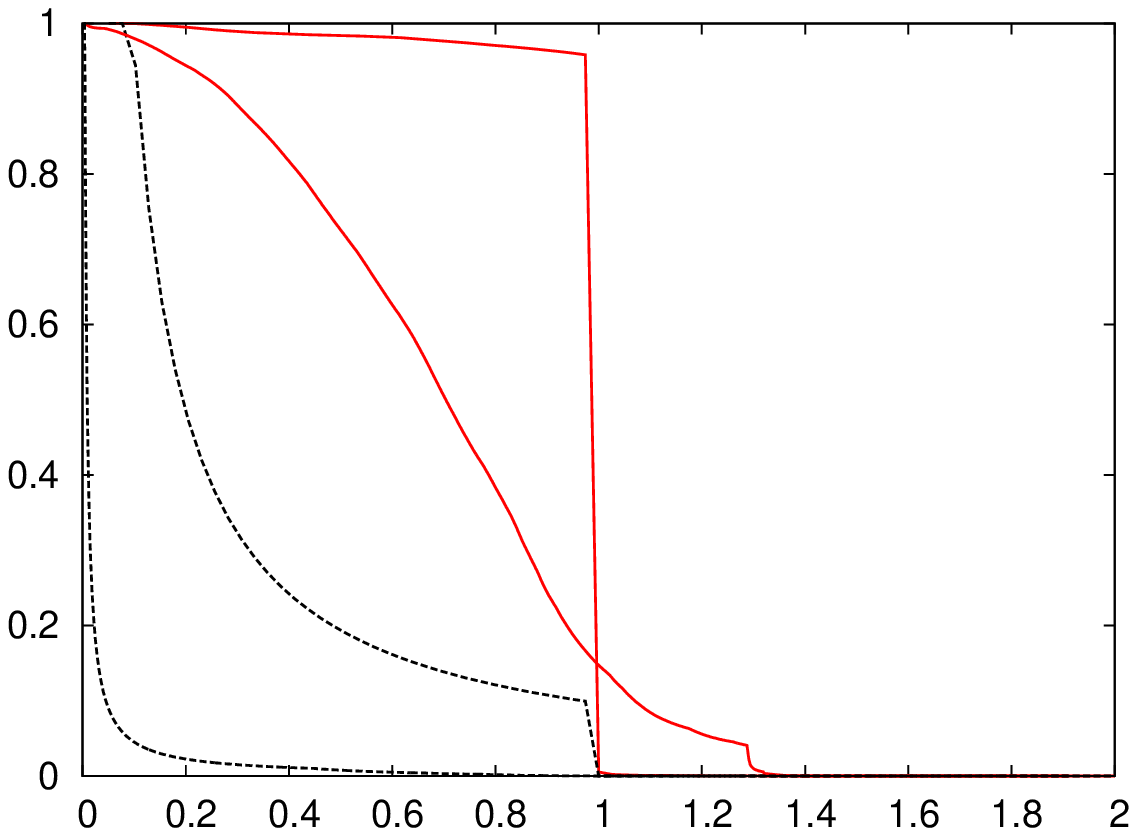}}
\put(120,-5){$n$}
\put(-15,100){$\rho_c(n)$}
\put(300,50){$\Delta_{min}$}
\put(350,150){$\Delta_{avg}$}
\put(370,-5){$\rho/\rho_c$}
\put(310,80){$n=0.5$}
\put(390,130){$n=0.5$}
\end{picture}
\caption{Left: results for the critical line $\rho_c(n)$ from Minover$^+$ (red) and belief propagation for the supports (green). Details are explained in the text. Right: plot of the minimum support size  (black-dotted lines) and average support size (red-solid lines) versus $\rho/\rho_c$ for the ratios $n=0.5$, $2$.}
\label{figcritical}
\end{figure}

As a second example we revisit the Von Neumann model \eqref{eq:VN}. We are interested in calculating the value of the critical line  $\rho_c(n)$ (with $n=N/M$) above which it is not possible to find more solutions to the set of inequalities \cite{DeMartino2005}. To find this critical line using the self-consistency equations for the supports, we assume that the critical line occurs when at least one support becomes zero. Thus for a fixed value of the ratio $n=N/M$ we increase the value of the global growth rate $\rho$ until at least one of the supports becomes zero.\\
Numerical results for the critical line are reported in the left panel of figure \ref{figcritical} and compared with the MinOver$^+$ algorithm. Here we have generated Poissonian graphs both for factor and variables nodes.  Factor nodes have an in-degree and out-degree average of 3. Thus the variable nodes have the in- and out-average degree equal $3/n$. The sizes of the graphs have been generated as follows: for $n<=1$, $M=500$ and $N$ is varied while for $n>1$ $N=500$ and $M$ is varied. The results are average over 50 graphs. Although the cavity equations for the supports can cope with much larger graphs we keep their sizes  small so as to compare with the MinOver$^+$ algorithm.
\label{examplecritical}

\begin{figure}[h]
\begin{picture}(300,200)
\put(0,0){\includegraphics[width=9cm, height=7cm]{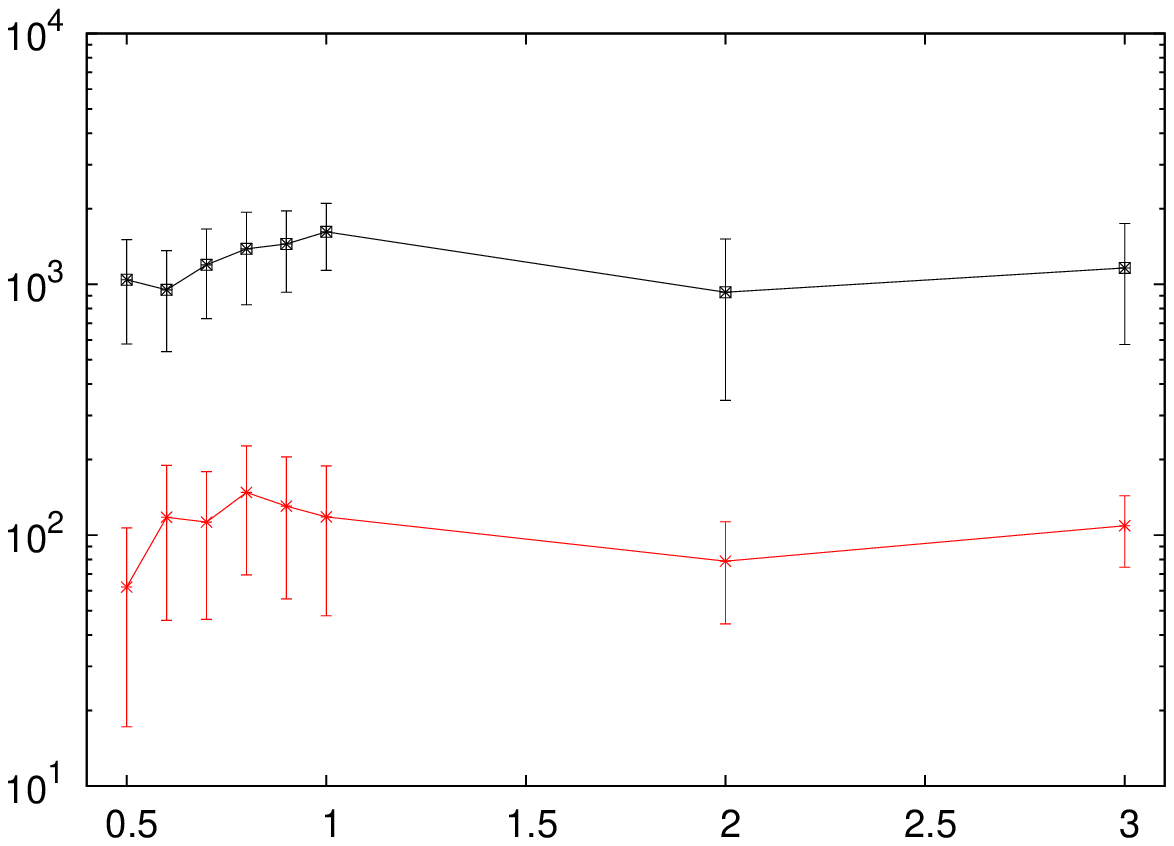}}
\put(120,-5){$n$}
\put(-5,100){$\bar{t}$}
\put(120,80){BP}
\put(150,150){MinOver$^+$}
\end{picture}
\caption{Average running time $\bar{t}$ versus $n$ for the belief propagation for the supports and the MinOver$^+$. As we can see the belief propagation is an order of magnitude faster.}
\label{fig3time}
\end{figure}

In the right panel of figure \ref{figcritical} we have plotted two possibles ``order parameters'' that could  be  used to locate the critical line $\rho_c$. These are the minimum support size $\Delta_{min}$ and the average support size $\Delta_{avg}$. It is interesting to notice that for $n\leq 1$ both parameters become zero at the critical line (as given by the MinOver$^+$ algorithm). This implies that at the critical line the volume of the polytope collapses to a point. Remarkably, for $n\geq 1$ we have  that $\Delta_{min}=0$ and $\Delta_{avg}>0$ at the critical line, indicating that the polytope has collapsed in one or more dimensions, but it is still dimensionfull in a subspace.\\
In figure \ref{fig3time} we also report the average running time $\bar{t}$ corresponding to figure \ref{figcritical} for both the belief propagation of the supports and the MinOver$^+$ algorithm. As it can be clearly observed, belief propagation is faster than MinOver$^+$ by an order of magnitude. Besides, the average running time $\bar{t}$ grows with $n=N/M$ for $n<1$, while remains constant for $n>1$. This is not unsurprising considering how the networks are generated (see description in the text above).

\subsection{A toy example for Weighted Population Dynamics in the Instance}
\label{example3g}
\begin{figure}[h]
\begin{picture}(500,200)
\put(0,0){\includegraphics[width=17cm, height=6.5cm]{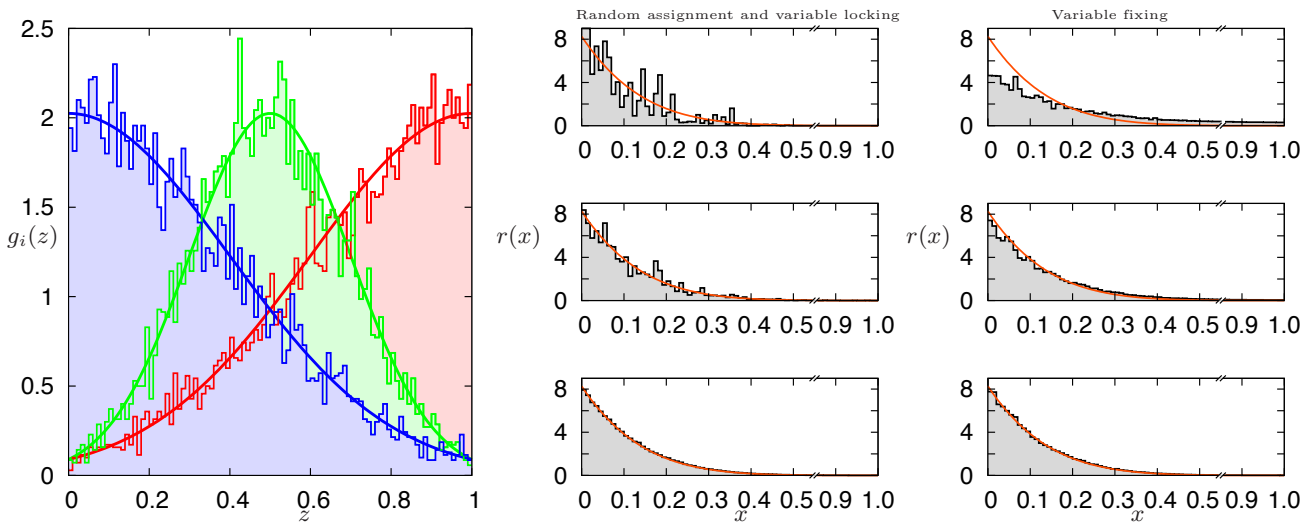}}
\put(210,185){\text{{\tiny Random assignment and variable locking}}}
\put(390,185){\text{{\tiny Variable fixing}}}
\put(180,100){$r(x)$}\put(335,100){$r(x)$}
\put(270,-5){$x$}\put(420,-5){$x$}
\put(-5,100){$g_i(z)$}\put(105,-5){$z$}
\end{picture}
\caption{In this figure we report the results of the weighted population dynamics in the instance applied to the toy example reported in \ref{example3g}. We have taken 3 Gaussian distributions (left panel) with means $0$, $1/2$, and $1$ and variances $\sigma^2=0.4,0.2$, and $0.4$. The centre panel corresponds to estimating $r(x)$ by WPDI using the method of random assignment \& variable locking using a population size of $\mathcal{N}=7\cdot 10^2$, $7\cdot 10^3$, and $7\cdot 10^5$ (top to bottom). The right panel corresponds to WPDI using the variable fixing method for a population of $\mathcal{N}=8\cdot 10^2$ and using averaged weighted using sizes $T=15,60,300$ (top to bottom). The orange solid line in the centre and right panels corresponds to the exact result for $r(x)$}
\label{fig3}
\end{figure}

In this section we illustrate \textit{random assignment \& variable locking} and \textit{variable fixing} in a particular case of the example  \eqref{example2} discussed before. As we want to illustrate the method to estimate the integral rather than its use to find the marginals, we consider here the the marginals to be known Gaussian distribution $g_i(y;m_i,\sigma^2_i)\equiv g_i(y)$  normalised on the interval $[0,1]$ and with mean value $m_i$ and variance $\sigma^2_i$.
\beeqn{
r(x)&=\frac{1}{r}I^2(x)\,,\quad I(x)=\int_{[0,1]^3}\left[\prod_{i=1}^3dy_ig_i(y_i)\right]\delta\left(x+\sum_{i=1}^3y_i-1\right)
}
where we also assume that $x\in[0,1]$. Although this a pedagogical example, it may be worth it, for the sake of clarity -but with the risk of being a bit repetitive- to discuss it in some detail. We describe here how the method of \textit{random assignment \& variable locking} looks like on a pen-and-paper calculation. Firstly, for one of the multiple integrals $I(x)$, we note that due to the assignment $x\leftarrow1-\sum_{i=1}^3y_i$, and the fact that $x,y_1,y_2,y_3\in[0,1]$, we have that from the integration region $[0,1]^3$ in the $(y_1,y_2,y_3)$-space, the region that actually contributes to the integral is the one below the plane $y_3=1-y_1-y_2$. Choosing an order of integration we write
\beeqn{
I(x)&=\int_{0}^1 dy_1g_1(y_1)\int_{0}^{1-y_1} dy_2 g_2(y_2|y_1)\int_{0}^{1-y_1-y_2} dy_3 g_3(y_3|y_1,y_2) w_2(y_1)w_3(y_1,y_2)\delta\left(x+\sum_{i=1}^3y_i-1\right)\\
}
where we have also reweighted the pdfs $g_2$ and $g_3$ so that is it normalised in the new integration intervals $[0,1-y_1]$ and $[0,1-y_1-y_2]$ respectively, with weights
\beeqn{
w_2(y_1)&=\int_{0}^{1-y_1} dy_2 g(y_2)=Q(y_1|m_2,\sigma_2^2)\,,\quad w_3(y_1,y_2)=Q(y_1+y_2|m_3,\sigma_3^2)
}
where we have defined the function 
\begin{equation*}
Q(x|a,b)\equiv\frac{\erf\left(\frac{a}{\sqrt{2b}}\right)-\erf\left(\frac{-1+a+x}{\sqrt{2b}}\right)}{\erf\left(\frac{a}{\sqrt{2b}}\right)-\erf\left(\frac{-1+a}{\sqrt{2b}}\right)}
\end{equation*}
and with  $g_2(y_2|y_1)=\frac{g_2(y_2)}{w_2(y_1)}$, $g_2(y_3|y_1,y_2)=\frac{g_3(y_3)}{w_3(y_1,y_2)}$. Then, to estimate the integral $I(x)$ and assign a value of $x$ we must implement the following steps:
\begin{enumerate}
\item Draw a Gaussian random variable $y_1$ with mean $m_1$ and variance $\sigma_1^2$ in the interval $[0,1]$, draw a Gaussian random variable $y_2$ with mean $m_2$ and variance $\sigma_2^2$ in the interval $[0,1-y_1]$, and draw Gaussian random variable $y_3$ with mean $m_3$ and variance $\sigma_3^2$ in the interval $[0,1-y_1-y_2]$,
\item Assign $x$ the value $x\leftarrow x^\star=1-\sum_{i=1}^3y_i$
\item Calculate the weight $w_2(y_1)w_3(y_1,y_2)$
\end{enumerate}
Now for the second $I(x)$ in our example, the value of $x$ is now locked at $x^\star$.  After restricting the integral to the region with non-zero value and using the Dirac delta to integrate over $z_3$ we can write
\beeqn{
I(x^\star)&=\int_{0}^{1-x^\star} dz_1g_1(z_1)\int_{0}^{1-z_1-x^\star} dz_2 g_2(z_2|z_1,x^\star) g_3[z_3(z_1,z_2,x^\star)] w_1(x^\star)w_2(y_1,x^\star)
}
with $z_3(z_1,z_2,x^\star)=1-x^\star-z_1-z_2$, and with the weights with the same definition as before. From here we see that to estimate $I(x^\star)$ by Monte Carlo we can do the following:
\begin{enumerate}
\item Draw a Gaussian random variable $z_1$ with mean $m_1$ and variance $\sigma_1^2$ on the interval $[0,1-x^\star]$ and draw a Gaussian random variable $z_2$ with mean $m_2$ and variance $\sigma_2^2$ on the interval $[0,1-x^\star-z_1]$
\item Caculate the weight $ g_3[z_3(z_1,z_2,x^\star)] w_1(x^\star)w_2(y_1,x^\star)$.
\end{enumerate}
From here the total weight is  $\omega=w_2(y_1)w_3(y_1,y_2)g_3[z_3(z_1,z_2,x^\star)] w_1(x^\star)w_2(y_1,x^\star)$ at $x=x^\star$. Repeating the process $\mathcal{N}$ times we obtain a population of pairs $\{(x_\alpha^\star,\omega_\alpha)\}_{\alpha=1}^{\mathcal{N}}$ from which to construct $r(x)$.\\
To implement the numerics we have chosen $m_1=0$, $m_2=1/2$, and  $m_3=1$ while for the standard deviations we have taken $\sigma_1=\sqrt{0.4}$, $\sigma_2=\sqrt{0.2}$, and  $\sigma_3=\sqrt{0.4}$ and we reported the results in figure \ref{fig3}, with a more detailed explanation in its caption.

\subsection{Random graph and Comparison with Hit-and-Run algorithm}
\begin{figure}[h]
\begin{center}
\includegraphics[width=12cm, height=16cm, angle=-90]{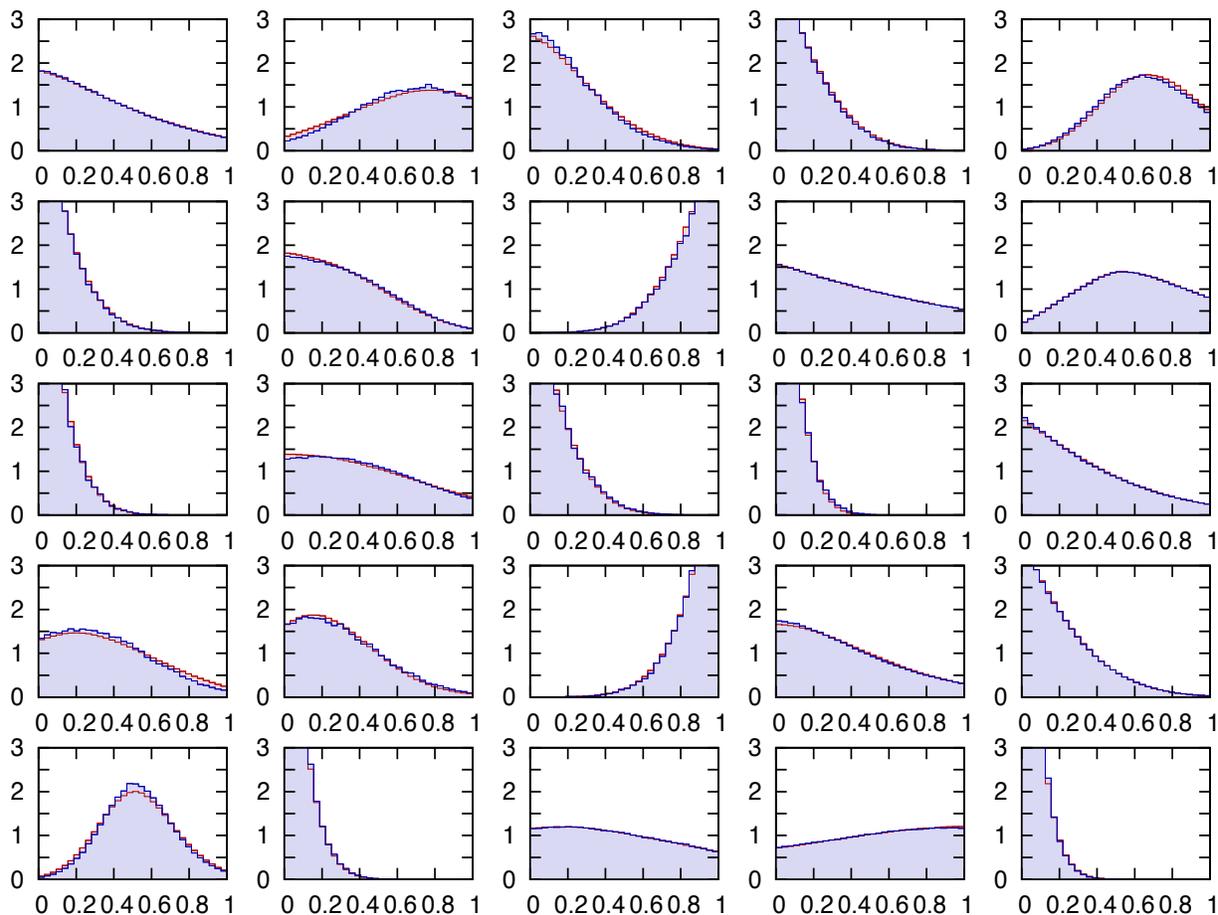}
\end{center}
\caption{Results for the marginals $P_i(x_i)$ for $i=1,\ldots,25$ (left to right and top to bottom) of the WPDI algorithm with variable fixing (blue) compared with estimates for the marginals using the hit-and-run algorithm (red)}
\label{fig4}
\end{figure}

Finally our last example corresponds to a random network of $N=25$ variable nodes and $M=10$ factor nodes. Although small, this size can be found in actual applications in metabolic networks as, for instance, human red blood cells \cite{DeMartino2010b,DeMartino2011}. Here we have used WPDI with variable fixing with a population size of $\mathcal{N}=10^3$. For the averaged weights we use a time window of $T=10^2$ weights for the cavity marginals and $T=10^6$ for the real marginals. Besides, for the final marginals a population is not used due to the simple fact that then we need to plot them, so we construct directly the histogram by fixing the value of $x$ at the midpoint at each bin. Results for each marginal $P_i(x_i)$ for $i=1,\ldots,25$ are presented in figure \ref{fig4} and we  have compared our findings with results obtained with the Hit-and-Run algorithm  (for implementation see appendix \ref{app:har}). As one can see the agreement is  excellent considering that the network is small and therefore loopy and our equations rely on the Bethe approximation.

\section{Conclusions}
\label{sect:con}
As we have discussed, many problems arising from diverse research topics may be mathematically recast as properties related to the volume of polytopes. Methods to estimate these properties can be roughly divided into mathematical and Monte Carlo approaches. In both cases, the approach used  may not be  efficient enough to treat polytopes of practical relevance (e.g. large metabolic networks). In this paper we have shown that for diluted random polytopes, a novel weighted belief propagation algorithm can be used to calculate efficient single-site marginals. We have discussed several options  of implementing this algorithm, devoting more lines to explain in detail the WPDI with its two versions:  {\em random assignment \& variable locking} and {\em variable fixing}. Importantly, we also pointed out that it is possible to write down self-consistency equations for the supports of the marginals which, apart of its use in implementing WPDI efficiently, can be employed  to obtain relevant information about the system with minor effort (e.g. to estimate the ranges of reaction rates allowed by stoichiometry; to evaluate the impact of a genetic mutation). Examples are used throughout this paper to illustrate,  to support, and to benchmark our novel theoretical findings and as so, we have kept most of them fairly simple. The attentive reader would have noticed that we have left some computational issues unattended. For instance we have not discussed in detail the fact that eq. \eqref{mtoP} is quadratic in the number of neigbours of $i$ and how this dependence can be linearised by adapting the nice trick in \cite{Braunstein2008}. Our main goal here is to provoke the reader with the novel ideas presented here rather than discussing ways to improve the algorithmic implementation. These issues are certainly important, but we believe they belong in another paper.\\
The work presented in this paper has several applications and extensions. First of all we aim to apply the weighted belief propagation to real metabolic networks, possibly starting from the human red blood cell, which is of size comparable to the example presented here, and extending the study to other reference metabolic networks (e.g. {\em E. coli}). Since we are able to tackle both FBA and the Von Neumann frameworks, we may use our tools for instance to compare these two approaches and to relate our results to the ones already published by using these two methods. Secondly, we plan to apply the weighted belief propagation to solve a metabolic network model where the direction of the reactions is not fixed and has to be determined by minimization of the Gibbs free energy. Finally, we plan to see how to use the self-consistency equations of the supports to study for instance the behaviour of metabolic networks under perturbations.

\appendix
\section{Derivation of single-site marginals  using the cavity method}
\label{appendixcavity}
The first step in deriving the cavity equations \eqref{mtoP},\eqref{Ptom} for problem \eqref{eq:inequalities} is to start by calculating the joint marginal $P_\mu(x_{\partial \mu})$ of the set of variables $x_{\partial \mu}$ connected to the factor node $\mu$. By definition this is given by
\beeqn{
P_\mu(x_{\partial\mu})&=\frac{1}{V}\int d\bm{x}_{\backslash \partial \mu}\prod_{\nu}f_\nu(x_{\partial \nu})=\frac{1}{V}f_{\mu}(x_{\partial \mu})\int d\bm{x}_{\backslash \partial\mu}\prod_{\nu(\neq \mu)}f_\mu(x_{\partial\mu})\\
&=\frac{1}{V_{\mu}}f_{\mu}(x_{\partial\mu})P_\mu^{(\mu)}(x_{\partial \mu})\,,\quad\quad P_\mu^{(\mu)}(x_{\partial \mu})=\frac{1}{V^{(\mu)}}\int d\bm{x}_{\backslash \partial\mu}\prod_{\nu(\neq \mu)}f_\mu(x_{\partial\mu})\,,
}
with $f_\mu(x_{\partial\mu})=\Theta[h_\mu(x_{\partial\mu})]$. Here, $P^{(\mu)}_\mu(x_{\partial \mu})$ is the joint marginal for the variables $x_{\partial \mu}$ in the absence of factor node $\mu$. If the graph is a tree or it is locally tree-like, the set of variables $x_{\partial \mu}$ in the absence of $\mu$ are mostly uncorrelated, and we can confidently write $P^{(\mu)}_\mu(x_{\partial \mu})=\prod_{\ell\in\partial\mu} P_{\ell}^{(\mu)}(x_\ell)$.\\
Let us move on to find an expression for the single site marginals $P_i(x_i)$:
\beeqn{
P_i(x_i)&=\frac{1}{V}\int d\bm{x}_{\backslash i}\prod_{\mu}f_\mu(x_{\partial\mu})\\
&=\frac{1}{V}\int d\bm{x}_{\backslash  i}\prod_{\mu\in \partial i}f_\mu(x_{\partial \mu})\prod_{\mu\not\in \partial i}f_\mu(x_{\partial\mu})\\
&=\frac{1}{V}\int dx_{(\partial\mu\ni i)\backslash i}\prod_{\mu\in \partial i}f_\mu(x_{\partial\mu})\int d\bm{x}_{\partial\mu\not\ni i}\prod_{\mu\not\in\partial  i}f_\mu(x_{\partial\mu})\,.
}
Here the notation $\bm{x}_{(\partial\mu\ni i)\setminus i}$ stands for the collection of variables nodes which share a factor node with $i$ with the exception of the node $i$ itself. The important point to note here is that $\int d\bm{x}_{\partial \mu\not\ni  i}\prod_{\mu\not\in \partial i}f_\mu(x_{\partial\mu})\propto P^{(\mu\in\partial i)}_{\mu\in\partial i}(x_{(\partial\mu\ni i)\backslash i})$. As variable node $i$ is absent in this marginalisation and as we are considering locally tree-like graphs we can confidently write $P^{(\mu\in\partial i)}_{\mu\in\partial i}(x_{(\partial\mu\ni i)\backslash i})=\prod_{\mu\in\partial i}P_\mu^{(\mu)}(x_{\partial \mu\setminus i})$. All in all we have the following expression:
\beeqn{
P_i(x_i)&=\frac{1}{V_i}\prod_{\mu\in \partial i}\int dx_{\partial \mu\backslash i}f_\mu(x_{\partial \mu})P^{(\mu)}_{\mu}(x_{\partial\mu \backslash i})\,.
}
Recalling that the joint marginals $P_\mu^{(\mu)}$ themselves factorise we finally write
\beeqn{
P_i(x_i)&=\frac{1}{V_i}\prod_{\mu\in \partial i}\int dx_{\partial\mu\backslash i}f_\mu(x_{\partial\mu})\prod_{j\in\partial \mu\backslash i} P^{(\mu)}_{j}(x_j)\,.
}
Close equations for the cavity marginals  $P^{(\mu)}_{j}(x_j)$ can readily be written down by simply removoing one of the factor nodes in the neighbourhood of $i$, that is:
\beeqn{
P^{(\nu)}_i(x_i)&=\frac{1}{V^{(\nu)}_i}\prod_{\mu\in \partial i\setminus\nu}\int dx_{\partial\mu\backslash i}f_\mu(x_{\partial\mu})\prod_{j\in\partial \mu\backslash i} P^{(\mu)}_{j}(x_j)\,.
}
Notice that the functions $m_\mu^{(i)}(x_i)$ we introduced in the main text are the combinations of terms in from of the first product in the predecing equation:
\beeqn{
m_\mu^{(i)}(x_i)=\frac{1}{m_{\mu}^{(i)}}\int dx_{\partial\mu\backslash i}f_\mu(x_{\partial\mu})\prod_{j\in\partial \mu\backslash i} P^{(\mu)}_{j}(x_j)\,.
}
\section{The Hit-and-Run algorithm}
\label{app:har}
As we mentioned previously, the Hit-and-Run algorithm is a numerical method to sample directly and uniformly the volume of a polytope. This useful feature makes it particularly attractive for problems such as FBA \cite{Almaas2004}, but has the unfortunate drawback of being rather slow. Nevertheless, we find it useful in our case as a benchmark for our message-passing algorithm.\\
We explain the Hit-and-Run algorithm by following the nice work in \cite{Berbee1987}, simply adapting the notation. Consider the following set of inequalities
\begin{equation*}
(\bm{\xi}^\mu)^T\bm{x} \geq \gamma^\mu\,,\quad\quad \mu=1,\ldots,M
\end{equation*}
with $\bm{x}\in\mathbb{R}^N$ and $\norm{\bm{\xi}^\mu}=1$, $\forall \mu=1,\ldots,M$ the unit vectors normal to the $M$ hyper-planes encapsulating the polytope.  Leaving aside some mathematical technicalities and simply assuming the polytope is properly defined, the  algorithm samples the volume by starting at a given point $\bm{X}$ inside the polytope and moving along a random direction $\bm{v}$. To know how far to move one needs to calculate all the interections between all hyperplanes and the straight line passing through $\bm{X}$ in the $\bm{v}$ direction. Then the closest hyperplanes would be the furthest region within the polytope that can be reached. This procedure can be iterated randomly as follows:
\begin{enumerate}
\item[Step 0]  Find a point $\bm{X}^{(0)}$ interior to the polytope. Set $n=0$.
\item[Step 1] Generate a direction vector $\bm{v}^{(n)}$ from a uniform distribution.
\item[Step 2] Determine the intersections
\begin{equation*}
\begin{split}
\lambda_{\mu}&=\frac{(\bm{\xi}^\mu)^T \bm{X}^{(n)}-\gamma^\mu}{(\bm{\xi}^\mu)^T \bm{v}^{(n)}}\,,\quad \mu=1,\ldots,M\\
\lambda^{+}&=\min_{1\leq \mu\leq M}\{\lambda_\mu|\lambda_\mu>0\}\,,\quad\quad \lambda^{-}=\max_{1\leq \mu\leq M}\{\lambda_\mu|\lambda_\mu<0\},
\end{split}
\end{equation*}
where $\lambda^{\pm}$ allow to determine the closest hyperplanes to $\bm{X}^{(n)}$.
\item[Step 3] Generate a uniform random number $u\in[0,1]$ and set
\begin{equation*}
\bm{X}^{(n+1)}=\bm{X}^{(n)}+\left(\lambda^{-}+u(\lambda^{+}-\lambda^{-})\right)\bm{v}^{(n)}
\end{equation*}
\item[Step 4] Set $n\to n+1$ and go to Step 1 or stop if satisfied with the sampling.
\end{enumerate}
It was proven \cite{Smith1984,Berbee1987} that the statistics of the sequence $\{\bm{X}^{(n)}\}_{n=0}^\infty$ converges to the uniform distribution on the polytope.

\acknowledgments
IPC would like to thank Lenka Zdeborov{\'a} for pointing out some work done in the area of compressed sensing. FFC would like to thank the hospitality of the department of Mathematics at King's College London. FAM aknowledges European Union Grants PIRG-GA-2010-277166 and PIRG-GA-2010-268342. FFC aknowledges Spanish Government grant FIS2009-09508. We would also like to thank the referees from JSTAT to help us make this a better well-rounded paper.

\bibliographystyle{apsrev}
\bibliography{bibliography}

\end{document}